\documentclass[aps,prd,onecolumn,amssymb]{revtex4}
\usepackage{graphicx,bm,color}
\usepackage{amsmath}
\usepackage{amssymb}
\usepackage{amsfonts}
\usepackage{epsfig}
\newcommand{\be}{\begin{equation}}
\newcommand{\ee}{\end{equation}}
\newcommand{\bea}{\begin{eqnarray}}
\newcommand{\eea}{\end{eqnarray}}
\newcommand{\beaa}{\begin{eqnarray*}}
\newcommand{\eeaa}{\end{eqnarray*}}

\allowdisplaybreaks[4]

\begin{document}

\title{Source of black bounces in Rastall gravity}
\author{K. Atazadeh$^1$}\email{atazadeh@azaruniv.ac.ir}
\author{H. Hadi$^{1,2}$}\email{hamedhadi1388@gmail.com}
\affiliation{$^1$Department of Physics, Azarbaijan Shahid Madani University, Tabriz, 53714-161 Iran\\ $^2$Faculty of Physics, University of Tabriz, Tabriz 51666-16471, Iran}

\begin{abstract}
	In this study, we explore the black bounce solution in Rastall gravity and its potential source field, which can be described as a black hole or wormhole solution depending on certain parameters. We focus on the Bardeen-Type black bounce and Simpson-Visser solution and aim to identify an appropriate source field for these solutions. Our findings suggest that in Rastall gravity, a source for the black bounce solution with non-linear electromagnetic can be found. However, in the presence of a non-linear electromagnetic source, it is impossible to identify an appropriate source for the black bounce solution without a scalar field. We also investigate the energy conditions outside the event horizon for two types of black bounce solutions: Simpson-Visser and Bardeen. We find that these solutions do not satisfy the null energy condition, but we also reveal that Rastall gravity has more flexibility for maintaining some of the energy conditions by selecting an appropriate value for the Rastall parameter $\gamma$.
\end{abstract}
\maketitle

\section{Introduction}

General relativity (GR) is widely considered to be the most appropriate and comprehensive theory to explain the gravitational interaction \cite{1,2}. This theory has demonstrated remarkable success in addressing numerous problems and possesses the capability to predict novel phenomena \cite{3,4,5,6,7,8,9,10}. Notably, the equations formulated by Einstein yield intriguing solutions, including the Schwarzschild black hole. This black hole is characterized by its static and symmetric nature, devoid of both spin and charge \cite{11}. Additionally, general relativity offers solutions known as wormholes, which connect disparate points in space-time within the same universe or even between different universes through a tunnel \cite{13}. The properties of these wormhole solutions, including their traversability, resemblance to black holes, and stability, have been extensively explored in various scholarly references \cite{14,15,16,17,18,19,20,21,22,23,24,25,26,27,28,29,30,31,32,33,34,35,36,37}.

A nonlinear electromagnetic source for the Einstein equations has yielded solutions for black holes and wormholes. Another type of solution that arises from this source is referred to as regular black holes, which possess an event horizon but lack a singularity. The initial proposal for this particular solution was put forth by Bardeen\cite{Bardeen}. A notable characteristic of regular black holes is the deviation of photons from geodesics, as well as alterations in the thermodynamics of these solutions  \cite{40,41,42}. For further details regarding this solution, refer to  \cite{43,44,45,46,47,48,49,50,51,52,53,54,55}.

A novel variant of a regular solution, commonly referred to as black bounce, possesses the capability to transform into either a black hole or a wormhole, contingent upon the specific selection of its parameters. Among these intriguing solutions, the Simpson-Visser black bounce stands out as a noteworthy example \cite{SV}. This particular solution exhibits a throat located at $r=0$, while the area of its event horizon remains unaffected by the solution's parameters. For alternative models of black holes, one may consult references\cite{57,58,59,60}. Extensive investigations have been conducted on various properties of these models, without explicitly attributing them to the source of matter \cite{61,62,63,64,65,66,67,68,69,70,71,72,73,74,75,76,77,78,79}. However, it should be noted that in the GR context, nonlinear electromagnetic fields alone are insufficient in describing the matter source responsible for the existence of the black bounce. To identify a suitable source for this solution, the inclusion of the scalar and phantom fields becomes necessary, see \cite{main}.
In this work, we are motivated to study the phenomenon of the black bounce within the framework of Rastall gravity theory. Our objective is to identify a suitable source for this solution in the presence of a non-linear electromagnetic field. The Rastall theory proposes a unique approach by coupling the matter and geometry fields in a non-minimal manner \cite{rastall}. This coupling challenges the conventional energy-momentum conservation law in curved spacetime, as it allows for particle creation during the evolution of the cosmos \cite{r7,r8,r9,r10,r11}. Remarkably, observational data supports the Rastall theory in the cosmological context, leading to intriguing results \cite{r13,r14}. For a comprehensive analysis of the thermodynamic aspects of this model in a flat Friedmann-Lema\^{\i}tre-Robertson-Walker (FLRW) universe, readers can refer to  \cite{r16,r17}. Furthermore, the Rastall theory has also been applied to consider the G\"{o}del-type universe \cite{godel}.

The paper is sectionalized as follows. Firstly, we delve into the examination of black bounce in Rastall gravity, focusing on its general form in section 2. Subsequently, in sections 3 and 4, we explore the Simpson-Visser solution and the Bardeen-type black bounce within the framework of Rastall gravity, respectively. In section 5, we consider the analysis of the energy conditions associated with these solutions, and also we plot energy conditions versus $r$ and Rastall parameter, $\gamma$. Finally, we close the paper with conclusions in section 6.

\section{Rastall theory of gravity and black bounce solution}
In this section, we will examine the Rastall theory within the context of a black bounce background. The black bounce solution is described by the line element given by
\begin{equation}\label{metric}
	ds^{2}=f(r)dt^{2}-f(r)^{-1}dr^{2}-\sigma(r)^{2}(d\theta^{2}+\sin^{2}\theta d\phi^{2}),
\end{equation}

where the functions  $\sigma(r)$ and $f(r)$ depend on the radial coordinate $r$. To illustrate the definition of $\sigma(r)$ and $f(r)$, we can refer to equation  $(\ref{simpsonsigma})$ in the subsequent section, which presents the Simpson-Visser solution of a black bounce that will be discussed further. However, it is important to note that this solution does not satisfy the vacuum Einstein equation. To address this, one possible approach is to introduce a non-linear electromagnetic field as the source for the black bounce. Due to the non-linearity of the electromagnetic field, the energy-momentum tensor does not have the property  $T^{0}_{0}\not=T^{1}_{1}$. Consequently, it is not possible to find a source from the non-linear electromagnetic field that yields $T_{00}=k^{2}L(F)$ and $T_{11}=k^{2}L(F)$, where $L(F)$ is an arbitrary function of $F$  and  $F$  is defined as $F=F^{\mu\nu}F_{\mu\nu}/4$, with $F^{\mu\nu}$ representing the Maxwell-Faraday tensor.

In this study, we utilize the Rastall theory to determine an appropriate source for the Black bounce solution. Rastall gravity, as proposed by Rastall \cite{rastall}, modifies the Einstein field equations. In this modified theory, for a spacetime with a Ricci scalar $R$ supported by a source of  $T^{\mu\nu}$, the following equation holds: 

\begin{equation}
    T^{\mu\nu}\,_{;\mu}=\lambda R^{,\nu},
\end{equation}
here,  $\lambda$ is referred to as the Rastall parameter. The Rastall field equation can be expressed as: 
\begin{equation}
	G_{\mu\nu}+\gamma g_{\mu\nu}R=k^{2}T_{\mu\nu},
\end{equation}
where $\gamma=k^{2} \lambda$  and  $k^{2}$ represents the Rastall gravitational coupling constant. Alternatively, this equation can be written as: 
\begin{equation}\label{action}
	G_{\mu \nu}=k^{2}S_{\mu\nu},
\end{equation}
where the effective momentum tensor $S_{\mu\nu}$ is defined as:
\begin{equation}\label{s}
	S_{\mu\nu}=T_{\mu\nu}-\frac{\gamma T}{4\gamma -1}g_{\mu \nu},
\end{equation}
the stress-energy tensor $T_{\mu\nu}$ is given by:
\begin{equation}
    T_{\mu\nu}=2\epsilon\partial_{\mu} \phi \partial_{\nu}\phi-g_{\mu\nu}(\epsilon g^{\alpha\beta} \partial_{\alpha} \phi \partial_{\beta}\phi -V(\phi))+g_{\mu\nu}L(F)-L_{F}F^{\alpha}_{\mu}F_{\nu\alpha},
\end{equation}
where $T^{\phi}_{\mu\nu}$ and  $T^{EM}_{\mu\nu}$   represent the stress-energy tensors for the scalar field and electromagnetic field, respectively. These are defined as: 
\begin{equation}
  T^{\phi}_{\mu\nu}=2\epsilon\partial_{\mu} \phi \partial_{\nu}\phi-g_{\mu\nu}(\epsilon g^{\alpha\beta} \partial_{\alpha} \phi \partial_{\beta}\phi -V(\phi)),  
\end{equation}
\begin{equation}
  T^{EM}_{\mu\nu}= g_{\mu\nu}L(F)-L_{F}F^{\alpha}_{\mu}F_{\nu\alpha}. 
\end{equation}
The distinction between phantom and scalar fields is denoted by the values of $\epsilon$, where  $\epsilon=-1$ and $\epsilon=+1$, respectively \cite{bron}. The function $L(F)$ is arbitrary and $L_{F}$ is its partial derivative with respect to the electromagnetic scalar $F$, i.e.,  $L_{F}=\frac{\partial L(F)}{\partial F}$. The expression for $F$ is given by 
\begin{equation}
    F=F^{\mu\nu}F_{\mu\nu}/4
\end{equation}

It should be noted that the theory incorporates a scalar field alongside the non-linear electromagnetic field. The solution will be determined for both sources, as it becomes evident that a consistent theory for black bounce cannot be achieved without the inclusion of a scalar field, even within the framework of Rastall gravity, and when solely considering non-linear electromagnetic fields. Similar to general relativity with Einstein equations, the identification of a source for non-linear electromagnetic fields is unattainable without the involvement of scalar or phantom fields.

The focus of our consideration lies solely on solutions that possess magnetic charges. Consequently, the Maxwell-Faraday tensor will have only one non-zero component which is defined as:
\begin{equation}
    F_{23}=Q\sin{\theta}.
\end{equation}
Therefore, the electromagnetic scalar is given by:
\begin{equation}\label{F}
    F(r)=\frac{Q^{2}}{2\sigma^{4}}.
\end{equation}
The $S_{\mu\nu}$ components concerning the equation  $(\ref{s})$  can be expressed as
\begin{equation}
    S_{00}=-f(r)\frac{\sigma^{4}(V(\phi)+L(r)+\epsilon(2\gamma-1)f(r)(\phi^{\prime})^{2})-2Q^{2}\gamma L_{F}}{(4\gamma -1)\sigma^{4}},
\end{equation}
\begin{equation}
    S_{11}=-\frac{-\sigma^{4}(V(\phi)+L(r)+\epsilon(-6\gamma+1)f(r)(\phi^{\prime})^{2})+2Q^{2}\gamma L_{F}}{(4\gamma -1)f(r)\sigma^{4}},
\end{equation}
 \begin{equation}
     S_{22}=-\frac{Q^{2}(1-2\gamma)L_{F}-\sigma(r)^{4}(V(\phi)+L(r)+\epsilon(2\gamma-1)f(r)(\phi^{\prime})^{2})}{(4\gamma -1)\sigma^{2}},
 \end{equation}
 \begin{equation}
     S_{33}=-\sin^{2}\theta \frac{Q^{2}(1-2\gamma)L_{F}-\sigma(r)^{4}(V(\phi)+L(r)+\epsilon(2\gamma-1)f(r)(\phi^{\prime})^{2})}{(4\gamma -1)\sigma^{2}},
 \end{equation}
where $\phi^{\prime}=\frac{d\phi}{dr}$. The effective equation of motion with respect to  $(\ref{action})$  is 
\begin{equation}\label{effectivefield}
     R_{\mu\nu}-\frac{1}{2}g_{\mu\nu}R=k^{2}S_{\mu\nu}.
 \end{equation}
 
 Now we apply the equation $(\ref{s})$ and $T_{\mu\nu}=T^{\phi}_{\mu\nu} + T^{EM}_{\mu\nu}$ into equation $(\ref{effectivefield})$, the equations of motion are given by:
 
 \begin{eqnarray}\label{1}
    -\frac{\sigma (r) \left(f'(r) \sigma '(r)+2 f(r) \sigma ''(r)\right)+f(r) \sigma '(r)^2-1}{\sigma (r)^2}=~~~~~~~~~~~~~~~~~~~~~~~~~~~~~~~~~~~~~~~~~~~~~~~~~~~~~~~~~~~~~~\\\nonumber
     k^{2}\frac{\sigma (r)^4 \left((2 \gamma -1) \epsilon  f(r) \varphi '(r)^2-L(r)-V(\varphi (r))\right)+2 \gamma  Q^2 L_F(r)}{(4 \gamma -1) \sigma (r)^4},
 \end{eqnarray}
 \begin{eqnarray}\label{2}
    \frac{1-\sigma '(r) \left(\sigma (r) f'(r)+f(r) \sigma '(r)\right)}{\sigma (r)^2}=~~~~~~~~~~~~~~~~~~~~~~~~~~~~~~~~~~~~~~~~~~~~~~~~~~~~~~~~~\\\nonumber
     k^{2}\frac{\sigma (r)^4 \left((1-6 \gamma ) \epsilon  f(r) \varphi '(r)^2-L(r)-V(\varphi (r))\right)+2 \gamma  Q^2 L_F(r)}{(4 \gamma -1) \sigma (r)^4},
 \end{eqnarray}
 \begin{eqnarray}\label{3}
    -\frac{\sigma (r) f''(r)+2 f'(r) \sigma '(r)+2 f(r) \sigma ''(r)}{2 \sigma (r)}=~~~~~~~~~~~~~~~~~~~~~~~~~~~~~~~~~~~~~~~~~~~~~~~~~~~~~~~~~~~~~~\\\nonumber
     k^{2}\frac{\sigma (r)^4 \left((2 \gamma -1) \epsilon  f(r) \varphi '(r)^2-L(r)-V(\varphi (r))\right)+(1-2 \gamma ) Q^2 L_F(r)}{(4 \gamma -1) \sigma (r)^4},
 \end{eqnarray}

where $\sigma^{\prime}=\frac{d\sigma}{dr}$ and $f^{\prime}=\frac{df}{dr}$ .  The scalar field  can be derived by utilizing equations $(\ref{1})$ and $(\ref{2})$ 
 \begin{equation}\label{derivativephi}
     \frac{d\phi}{dr}=\sqrt{\frac{1}{\sigma k^{2}}\frac{d^{2}\sigma}{dr^{2}}},
 \end{equation}
 where we used $\epsilon =-1$ . When $\epsilon=-1$ is applied, a real scalar field is produced, but the null energy condition is violated as $\frac{1}{\sigma k^{2}}\frac{d^{2}\sigma}{dr^{2}} > 0$. It is important to note that this scalar field is only real for a phantom scalar field. The null energy condition holds as  $-2\frac{f}{\sigma k^{2}}\frac{d^{2}\sigma}{dr^{2}}\ge 0$ where $f>0$. However, if $\frac{1}{\sigma k^{2}}\frac{d^{2}\sigma}{dr^{2}} < 0$ the null energy condition is satisfied and a real scalar field is produced for  $\epsilon =1$.
 Therefore, in the context of Rastall gravity, when a scalar field and nonlinear electrodynamics are applied to produce a source for a black bounce solution, the scalar field must be phantom if $\frac{1}{\sigma }\frac{d^{2}\sigma}{dr^{2}} > 0$ leading to violation of null energy condition. The violation of the null energy condition is a prerequisite for any such solution to exist. 
  Consequently, in such scenarios, the value of $\epsilon$ is equal to $-1$.\\
 It is important to note that while the energy conditions apply to ordinary classical matter, they can be violated when dealing with quantized matter fields like the Casimir vacuum energy. Although quantum effects allow for localized violations of the energy conditions, there is a limit to how much they can be globally violated.
 In this regard, it is beneficial to formulate averaged versions of the energy conditions. One example is the averaged null energy condition, which states that the integral of $T_{\mu\nu}k^{\mu}k^{\nu}$ along a null geodesic $\vartheta$ must be non-negative. This implies that $\int_{\vartheta}T_{\mu\nu}k^{\mu}k^{\nu} d\lambda\ge 0$.
These averaged energy conditions play a significant role in the study of traversable wormholes. For further discussion on this topic, refer to Ref. \cite{13}.

 One can derive the non-linear electromagnetic quantities $L(r)$, $L_{F}(r)$ and scalar potential $V(\phi)$ by using the equations $(\ref{1})$,$(\ref{2})$ and $(\ref{3})$ in the following manner:

 \begin{eqnarray}\label{s1}
     L(r)=-[2\gamma -1]f\epsilon(\frac{d\phi}{dr})^{2}-\frac{f(1-2\gamma)}{k^{2}\sigma^{2}}\left(\frac{d\sigma}{dr}\right)^{2}\\\nonumber
     +\frac{4\gamma-1}{\sigma k^{2}}\frac{d\sigma}{dr}\frac{df}{dr}+2\gamma\frac{f}{\sigma k^{2}}\frac{d^{2}\sigma}{dr^{2}}+\frac{\gamma}{k^{2}}\frac{d^{2}f}{dr^{2}}+\frac{1-2\gamma }{k^{2}\sigma^{2}}-V(\phi),
 \end{eqnarray}
 \begin{equation}\label{s2}
     L_{F}(r)=+\frac{\sigma^{2}}{k^{2}Q^{2}}\left(\sigma f \frac{d^{2}\sigma}{dr^{2}}+\frac{1}{2}\sigma^{2}\frac{d^{2}f}{dr^{2}}-f\left(\frac{d \sigma}{dr}\right)^{2}+1\right)+\frac{2\epsilon f \sigma ^{4} }{Q^{2}}\left(\frac{d \phi}{dr}\right)^{2},
 \end{equation}
 \begin{equation}\label{s3}
     \frac{dV}{dr}=2\epsilon \frac{d\phi}{dr}\left(\frac{2f}{\sigma}\frac{d\sigma}{dr}\frac{d\phi}{dr}+\frac{df}{dr}\frac{d\phi}{dr}+f\frac{d^{2}\phi}{dr^{2}}\right).
 \end{equation}
It should be noted that, in the case of Rastall gravity and without the presence of magnetic charge $Q$, the functions $L(r)$ and $L_{F}(r)$ become zero, and thus no black bounce solution can be obtained with only a scalar field. From equations $(\ref{1})$, $(\ref{2})$ and $(\ref{3})$, with $L(r)$, $L_{F}(r)$ and $Q$ set to zero, we have three equations with two unknowns, $\varphi$ and $V(\varphi)$. Therefore, the solution is not unique.

The forthcoming sections will outline the determination of $f(r)$ and $\sigma (r)$, resulting in the derivation of equations $(\ref{s1})$, $(\ref{s2})$ and $(\ref{s3})$. The Simpson-Visser and Bardeen-Type black bounce solutions will be implemented in Rastall gravity.

 \section{Simpson-Visser solution in Rastall gravity}
The line element defining the black bounce solution in the Simpson-Visser model is as follows
\begin{equation}
	ds^{2}=f(r)dt^{2}-f(r)^{-1}dr^{2}-\sigma(r)^{2}(d\theta^{2}+\sin^{2}\theta d\phi^{2}),
\end{equation}
where
\begin{equation}\label{simpsonsigma}
	f(r)=1-\frac{2m}{\sqrt{r^{2}+Q^2}} ~~~and~~~ \sigma(r)=\sqrt{r^{2}+Q^2}.
\end{equation}

The parameter $Q$ can be interpreted as the magnetic charge, which plays a crucial role in preventing the occurrence of a singularity at $r=0$. This can be demonstrated by calculating the \textit{Kretschmann scalar}, which is a scalar invariant that measures the curvature of spacetime. The \textit{Kretschmann scalar} for this solution is given by
\begin{equation}
R_{\mu\nu\rho\sigma}R^{\mu\nu\rho\sigma}=\frac{m^2 \left(36 Q^4-48 Q^2 r^2+47 r^4\right)+32 m Q^2 \left(r^2-Q^2\right) \sqrt{Q^2+r^2}+12 Q^4 \left(Q^2+r^2\right)}{\left(Q^2+r^2\right)^5}
\end{equation}
As can be seen from the above expression, the Kretschmann scalar remains finite at $r=0$ as long as $Q \neq 0$. Therefore, the magnetic charge $Q$ acts as a regularizing factor that avoids the formation of a central singularity at $r=0$. This implies that the solution represents a black bounce, which is a nonsingular black hole and also has a wormhole solution.

However, this particular solution fails to meet the requirements of Einstein's equations in a vacuum. Consequently, the matter content is understood to be an anisotropic fluid in the following manner
 \begin{equation}
 	\rho=-\frac{Q^{2}(\sqrt{r^2+Q^{2}}-4m)}{8\pi(r^{2}+Q^{2})^{5/2}},
 \end{equation}
 \begin{equation}
 	p_{1}=-\frac{Q^{2}}{8\pi(r^{2}+Q^{2})^{2}},
 \end{equation}
 \begin{equation}
 	p_{2}=-\frac{Q^{2}(\sqrt{r^{2}+Q^{2}}-m)}{8\pi(r^{2}+Q^{2})^{\frac{5}{2}}},
 \end{equation}
the quantities $\rho$, $p_{1}$ and $p_{2}$  are defined as the components of stress-energy tensor which is given by
\begin{equation}
	T^{\mu}_{\nu}={\rm diag}[\rho, -p_{1},-p_{2},-p_{2}].
\end{equation}
Using equations $(\ref{derivativephi})$, $(\ref{simpsonsigma})$, $(\ref{s1})$, $(\ref{s2})$ and $(\ref{s3})$ we can derive the scalar field, the potential, and the electromagnetic quantities for this solution. The expression for $\frac{d\phi}{dr}$ is given by $\frac{d\phi}{dr}=\frac{Q}{k(r^{2}+Q^{2})}$. The other quantities can be written as follows
 \begin{equation}\label{phi}
     \phi (r)=\frac{1}{k} \arctan \left(\frac{r}{Q}\right),
 \end{equation}
 \begin{equation}\label{V}
     V(r)= \frac{4m Q^{2}}{5k^{2}(Q^{2}+r^{2})^{5/2}},
 \end{equation}
 \begin{equation}\label{L}
     L(r)=
     \frac{6mQ^{2}}{5k^{2}(r^{2}+Q^{2})^{5/2}}
     +\frac{\gamma}{k^{2}(r^{2}+Q^{2})^{5/2}}[2(r^{2}+Q^{2})^{3/2}-6mQ^{2}],
 \end{equation}
 \begin{equation}\label{LF}
     L_{F}(r)=\frac{3m}{k^{2}\sqrt{r^{2}+Q^{2}}}.
 \end{equation}
In the absence of $\gamma$, the solution coincides with that of GR when scalar and non-linear electromagnetic fields are present. To obtain explicit expressions for $V(\phi)$ and $L(F)$, we substitute $(\ref{phi})$ into $(\ref{V})$ and $(\ref{F})$  into $(\ref{L})$ which lead to 

 \begin{equation}
     V(\phi)=\frac{4m \cos^{5}(\phi k)}{5k^{2}|Q|^{3}},
 \end{equation}
 \begin{equation}
     L(F)=\frac{12(2)^{1/4}m F^{5/4}}{5k^{2}\sqrt{|Q|}}
     +\frac{\gamma (2F)^{5/4}}{k^{2}(|Q|)^{5/2}}[2(\frac{Q^{2}}{2F})^{3/4}-6mQ^{2}].
 \end{equation}
 
\begin{figure}
    \centering
    \includegraphics[width=0.5\linewidth]{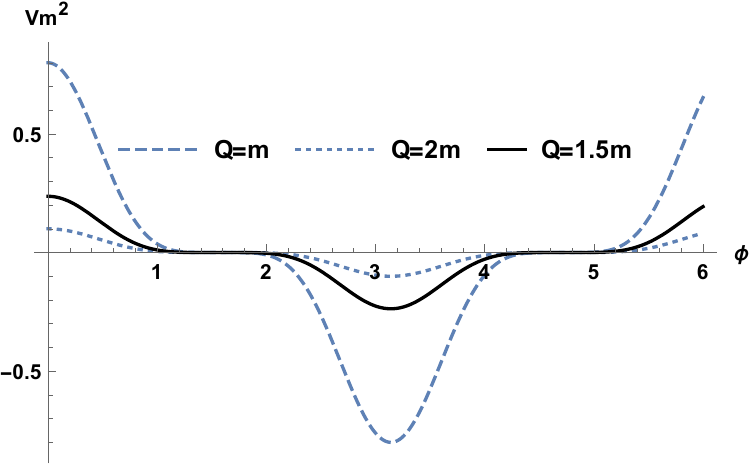}
    \caption{The scalar field potential, as a function of $\phi$, that produces the Simpson–Visser solution for different values of magnetic charge. }
    \label{fig:1}
\end{figure}
\begin{figure}
    \centering
    \includegraphics[width=0.5\linewidth]{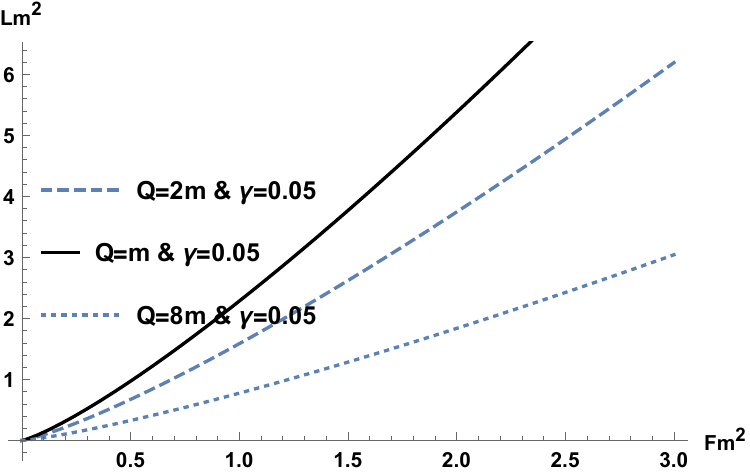}
    \caption{The electromagnetic Lagrangian, as a function of $F$,
 that generates the Simpson–Visser solution for different values of charge and a particular value of $\gamma$.  }
    \label{fig:2}
\end{figure}
\begin{figure}
    \centering
    \includegraphics[width=0.5\linewidth]{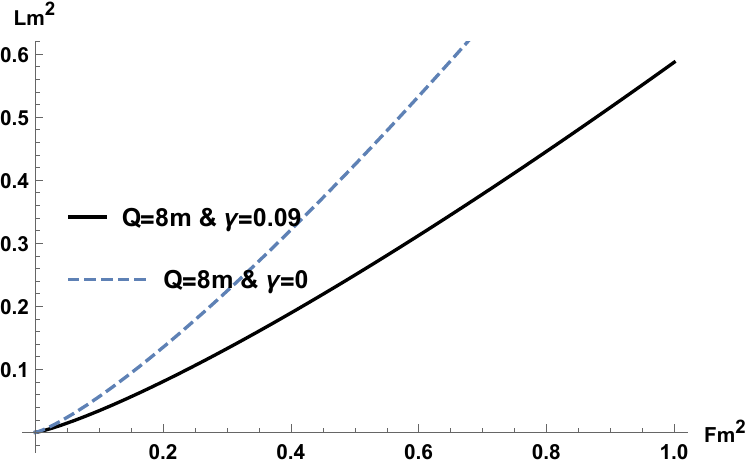}
    \caption{The electromagnetic Lagrangian, as a function of $F$,
 that generates the Simpson–Visser solution for a particular value of charge $Q=8m$ in the absence and presence of the Rastall parameter $\gamma$.}
    \label{fig:3}
\end{figure}

Figure$(\ref{fig:1})$ displays the potential $V(\phi)$ for various magnetic charge  $Q$ values with respect to $\phi$. It is observed that the amplitude of the potential decreases as $Q$ increases, and this characteristic is independent of the $\gamma$ parameter. Figure$(\ref{fig:2})$ illustrates the function $L(F)$ about the electromagnetic scalar. The figure shows that for a specific value of $\gamma$, an increase in $Q$ results in a decrease in the value of $L(F)$. The figure$(\ref{fig:3})$ for Simpson-Visser black bounce indicates that a larger $\gamma$ leads to a greater decrease in the value of $L(F)$.

\begin{figure}
    \centering
    \includegraphics[width=0.5\linewidth]{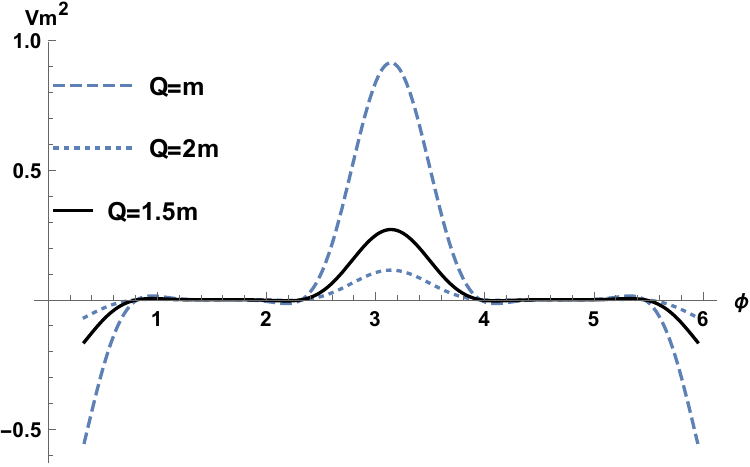}
    \caption{Behavior of the scalar field potential, as a function of $\phi$,
 that generates the Bardeen-type solution for different values of magnetic charge.}
    \label{fig:4}
\end{figure}

\begin{figure}
    \centering
    \includegraphics[width=0.5\linewidth]{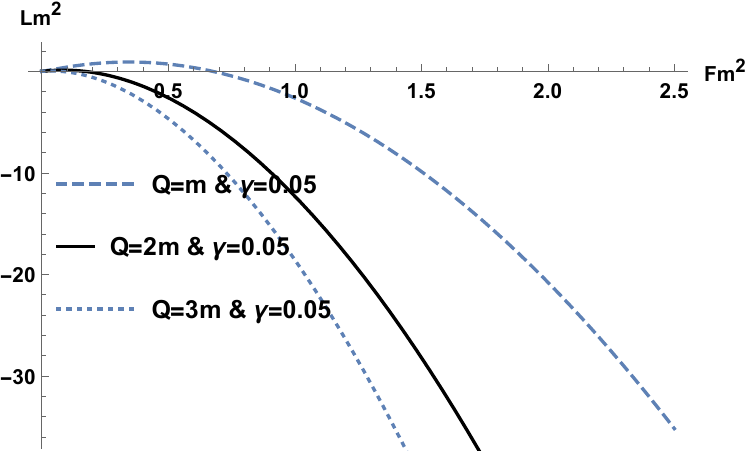}
    \caption{The electromagnetic Lagrangian, as a function of $F$,
 that produces the Bardeen-type solution for different values of charge and a particular value of $\gamma$. }
    \label{fig:5}
\end{figure}
\begin{figure}
    \centering
    \includegraphics[width=0.5\linewidth]{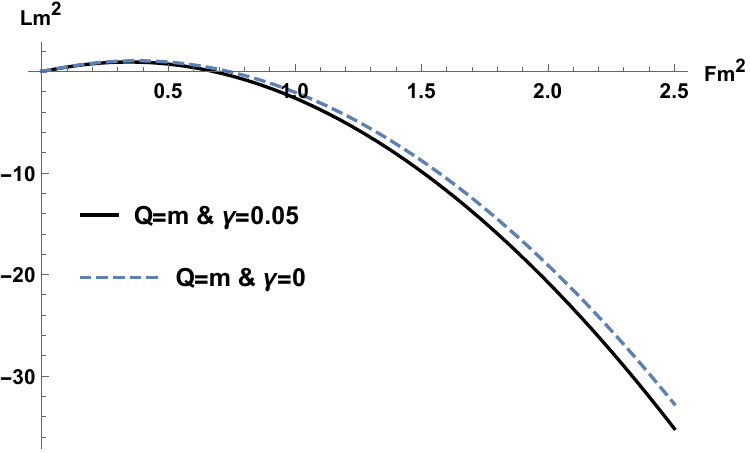}
    \caption{The electromagnetic Lagrangian, as a function of $F$,
 that generates the Bardeen-type solution for a particular value of charge  $Q=m$  when $\gamma$ is zero and $\gamma=0.05$ .}
    \label{fig:6}
\end{figure}

  \section{Bardeen-Type black bounce in Rastall theory}

In this section, we will examine the Bardeen-type black bounce in the Rastall model. The line element that describes this solution is provided by the metric  $(\ref{metric})$, along with the functions:
\begin{equation}\label{sigma2}
     f(r)=1-\frac{2mr^{2}}{(r^{2}+Q^{2})^{3/2}} ~~~~~~~ and~~~~~\sigma (r)=\sqrt{r^{2}+Q^{2}}.
 \end{equation}
As we discussed for the Simpson-Visser black bounce, the parameter $Q$, which we interpreted as the magnetic charge, prevents the occurrence of a central singularity at $r=0$. Similarly, for the Bardeen-Type solution, we have a parameter $Q$ that acts as a regularizing factor that avoids the formation of a singularity at $r=0$. The Kretschmann scalar for this solution is given by
\begin{eqnarray}
 R_{\mu\nu\rho\sigma}R^{\mu\nu\rho\sigma}=~~~~~~~~~~~~~~~~~~~~~~~~~~~~~~~~~~~~~~~~~~~~~~~~~~~~~~~~~~~~~~~~~~~~~~~~~~~~~~~~~~~~~~~~~~~~~~~~~~~~~~~~~~~~~~~~~~~~~~~~~~~~~~~~\\\nonumber
 \frac{m^2 \left(16 Q^8-176 Q^6 r^2+672 Q^4 r^4-268 Q^2 r^6+47 r^8\right)-32 m Q^2 r^2 \sqrt{Q^2+r^2} \left(2 Q^4+Q^2 r^2-r^4\right)+12 Q^4 \left(Q^2+r^2\right)^3}{\left(Q^2+r^2\right)^7}   
\end{eqnarray}
As can be seen from the above expression, the Kretschmann scalar remains finite at $r=0$ as long as $Q \neq 0$. Therefore, the parameter $Q$ plays a similar role as in the Simpson-Visser case, and the solution represents a black bounce.

By utilizing equations $(\ref{derivativephi})$, $(\ref{sigma2})$, $(\ref{s1})$, $(\ref{s2})$ and $(\ref{s3})$, we can derive the scalar field, the potential, and the electromagnetic quantities associated with this solution.  The expression for $\frac{d\phi}{dr}$ is given by $\frac{d\phi}{dr}=\frac{Q}{k(r^{2}+Q^{2})}$. The remaining quantities can be expressed as follows
 \begin{equation}\label{V2}
     V(r)=\frac{4m(7Q^{2}r^{2}-8Q^{4})}{35k^{2}(Q^{2}+r^{2})^{7/2}},
 \end{equation}
 \begin{equation}\label{L2}
     L(r)=\frac{2mQ^{2}(16Q^{2}+91r^{2})}{35k^{2}(Q^{2}+r^{2})^{7/2}}
     - \frac{2\gamma mQ^{2}}{k^{2}(r^{2}+Q^{2})^{7/2}}(r^{2}+2Q^{2}),
 \end{equation}
 \begin{equation}\label{LF2}
     L_{F}(r)=\frac{m(13r^{2}-2Q^{2})}{k^{2}(Q^{2}+r^{2})^{3/2}}.
 \end{equation}

In order to derive explicit equations for $V(\phi)$ and $L(F)$, the substitution of  $(\ref{phi})$ into $(\ref{V2})$ and $(\ref{F})$ into $(\ref{L2})$ is performed, resulting in the following expressions

 \begin{equation}
V(\phi)=\frac{4m \cos^{2}(\phi k)(7\sin^{2}(\phi k)-8\cos^{2}(\phi k))}{35k^{2}|Q^{3}|},
 \end{equation}
 \begin{equation}
     L(F)= 
     \frac{4(2)^{1/4}F^{5/4}m(91-75(2F)^{1/2}Q)}{35k^{2}|Q|^{1/2}}
     -\frac{2\gamma m(2F)^{5/4}}{k^{2}|Q|^{1/2}}[1+Q\sqrt{2F}].
 \end{equation}
The potential $V(\phi)$ as a function of $\phi$ for different values of magnetic charge $Q$ is shown in Figure$(\ref{fig:4})$. The figure reveals that the potential has a lower amplitude for higher values of $Q$, regardless of the $\gamma$ parameter. The function $L(F)$ versus the electromagnetic scalar is depicted in Figure$(\ref{fig:5})$. The figure demonstrates that for a fixed value of $\gamma$, a higher value of $Q$ corresponds to a lower value of $L(F)$. The figure$(\ref{fig:6})$ for Bardeen-Type black bounce indicates that the value of $L(F)$ decreases more rapidly for larger values of $\gamma$.

 \section{Energy conditions}
The energy condition typically arises when investigating the Raychaudhuri equation, which is expressed as:
\begin{equation}
\frac{d \theta}{d\tau}=-\frac{1}{2}\theta^{2}-\mathcal{\sigma}_{\mu\nu}\mathcal{\sigma}^{\mu\nu}+\omega_{\mu\nu}\omega^{\mu\nu}-R_{\mu\nu}k^{\mu}k^{\nu},
\end{equation}
where, $\theta$, $\mathcal{\sigma}_{\mu\nu}$, and $\omega_{\mu\nu}$ represent the expansion, shear, and rotation, respectively. The equation is satisfied when $R_{\mu\nu}k^{\mu}k^{\nu}$ is applied. For gravity to exhibit attractive behavior, it is necessary that $\frac{d\theta}{d\tau}<0$.

In the context of general relativity, utilizing the inequality equation, one can reformulate the condition as an energy-momentum tensor denoted by $T_{\mu\nu}k^{\mu}k^{\nu} \ge 0$. However, in alternative theories of gravity, it becomes crucial to transform the equation into a form resembling Einstein's field equation (in its effective form) by incorporating a modified stress-energy tensor.

In Rastall gravity, where the field equation can be written as $R_{\mu\nu}-\frac{1}{2}Rg_{\mu\nu}=k^{2}S_{\mu\nu}$, thus according to the Einstein field equations one can employ the effective energy-momentum tensor $S_{\mu\nu}$ in the right side of field equations instead of the standard energy-momentum tensor $T_{\mu\nu}$ to establish the energy conditions.

The Rastall gravity and its energy conditions, specifically the null energy condition, are thoroughly analyzed in this section. The violation of this condition is observed in the case of a black bounce in general relativity for a phantom field with $\epsilon=-1$, which is crucial for the existence of a real scalar field. Additionally, the section provides a comprehensive examination of all energy conditions for both scalar and non-linear electromagnetic fields, to the Rastall stress-energy tensor $S^{\mu}_{\nu}$. The energy conditions being considered are as follows

\begin{figure}[!tbp]
  \centering
  \begin{minipage}[b]{0.49\textwidth}
    \includegraphics[width=\textwidth]{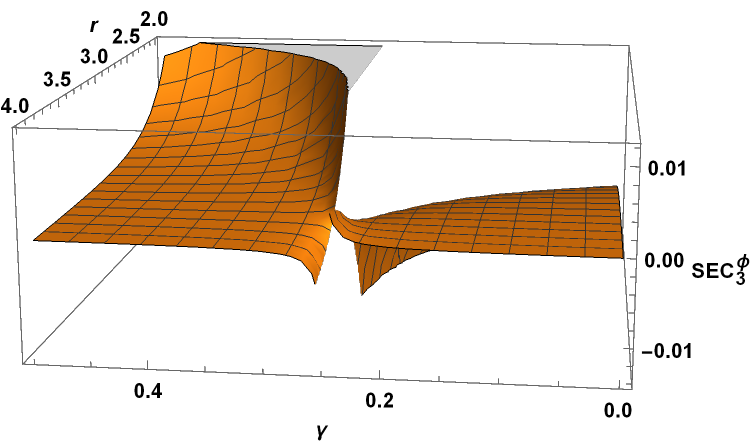}
    \caption{ ${\rm SEC}_{3}^{\phi}$ for different values of parameter $\gamma$ and $r$ outside the horizon for a particular value of $m=1$ and $Q=0.5$ in the Simpson-Visser model.}
  \end{minipage}
  \hfill
  \begin{minipage}[b]{0.49\textwidth}
    \includegraphics[width=\textwidth]{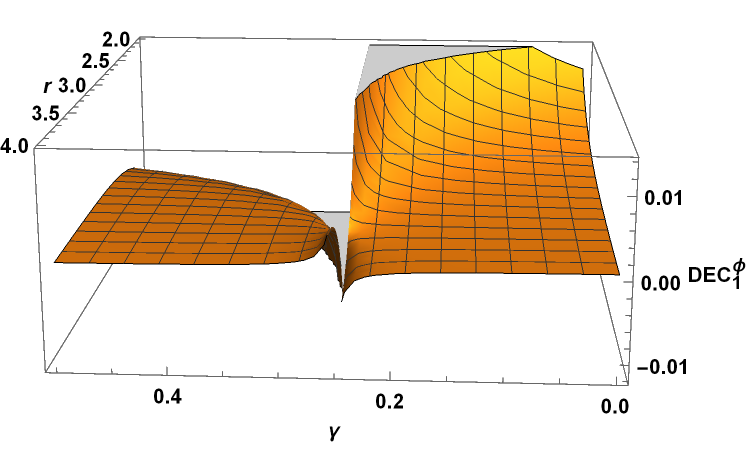}
    \caption{ ${\rm DEC}_{1}^{\phi}$ for different values of parameter $\gamma$ and $r$ outside the horizon for a particular value of $m=1$ and $Q=0.5$ in the Simpson-Visser model.}
  \end{minipage}
  \end{figure}

\begin{equation}\label{e1}
     {\rm NEC}_{1,2}^{\phi, EM}= {\rm WEC}_{1,2}^{\phi, EM}= {\rm SEC}_{1,2}^{\phi, EM}\Longleftrightarrow \rho^{\phi, EM}+p_{1,2}^{\phi, EM}\ge 0.
 \end{equation}
 The null energy condition $({\rm NEC})$ can be geometrically represented as $R_{\mu\nu}k^{\mu}k^{\nu} \ge 0$ for any null vector $k^{\mu}$, while its physical meaning is translated as $S_{\mu\nu}k^{\mu}k^{\nu} \ge 0$. An effective form of ${\rm NEC}$ can be defined as $\rho + p_{a} \ge 0$ for each value of $a$. It is worth noting that this effective form is the same as the effective form of weak energy condition $({\rm WEC})$ and strong energy condition $({\rm SEC})$, but their geometric and physical meanings are not equivalent. The geometric meaning of ${\rm WEC}$ for any timelike vector $\zeta^{\mu}$ is obtained by $G_{\mu\nu}\zeta^{\mu}\zeta^{\nu}\ge 0$. On the other hand, the geometric form of ${\rm SEC}$ is the same as ${\rm NEC}$, but instead of a null vector $k^{\mu}$, a timelike vector is applied as $R_{\mu\nu}\zeta^{\mu}\zeta^{\nu} \ge 0$. However, its physical meaning is entirely different and is given by $(T_{\mu\nu}-\frac{1}{2}Tg_{\mu\nu})\zeta^{\mu}\zeta^{\nu}\ge 0$
 \begin{figure}
 \centering
\begin{minipage}[b]{0.49\textwidth}
    \includegraphics[width=\textwidth]{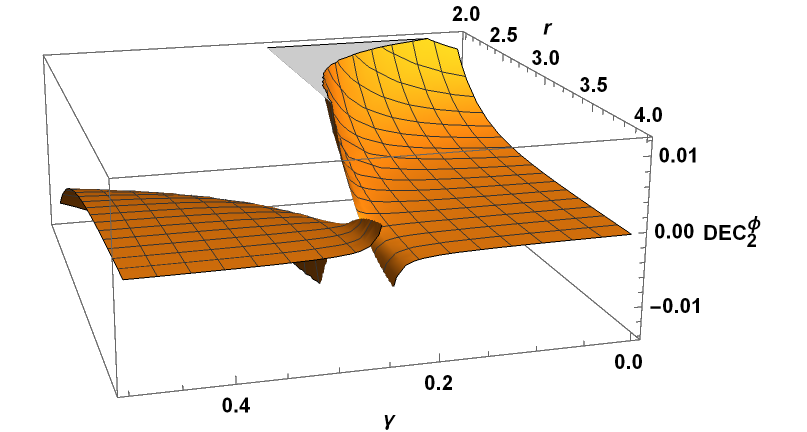}
    \caption{ ${\rm DEC}_{2}^{\phi}$ for different values of parameter $\gamma$ and $r$ outside the horizon for a particular value of $m=1$ and $Q=0.5$ in the Simpson-Visser model.}
  \end{minipage}
  \hfill
  \begin{minipage}[b]{0.49\textwidth}
    \includegraphics[width=\textwidth]{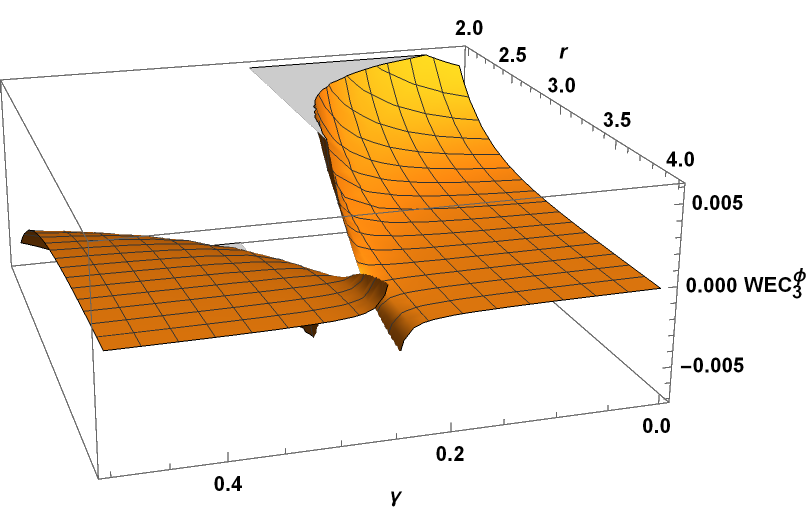}
    \caption{ ${\rm WEC}_{3}^{\phi}$ for different values of parameter $\gamma$ and $r$ outside the horizon for a particular value of $m=1$ and $Q=0.5$ in the Simpson-Visser model.}
  \end{minipage}
\end{figure}

\begin{figure}[!tbp]
  \centering
  \begin{minipage}[b]{0.49\textwidth}
    \includegraphics[width=\textwidth]{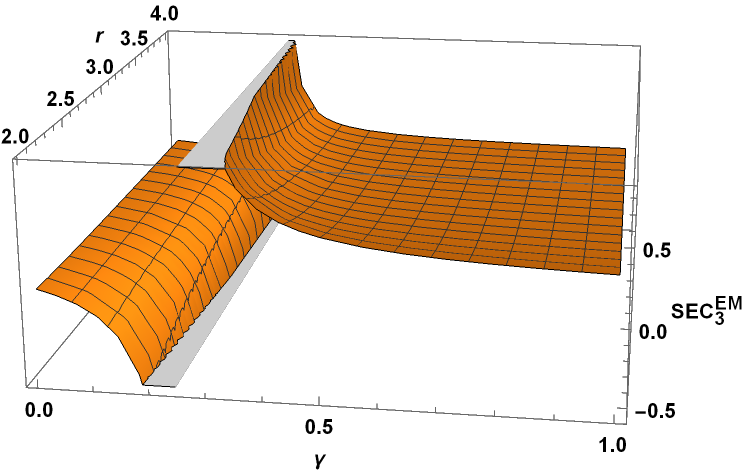}
    \caption{ ${\rm SEC}_{3}^{EM}$ for different values of parameter $\gamma$ and $r$ outside the horizon for a particular value of $m=1$ and $Q=0.5$ in the Simpson-Visser model.}
  \end{minipage}
  \hfill
  \begin{minipage}[b]{0.49\textwidth}
    \includegraphics[width=\textwidth]{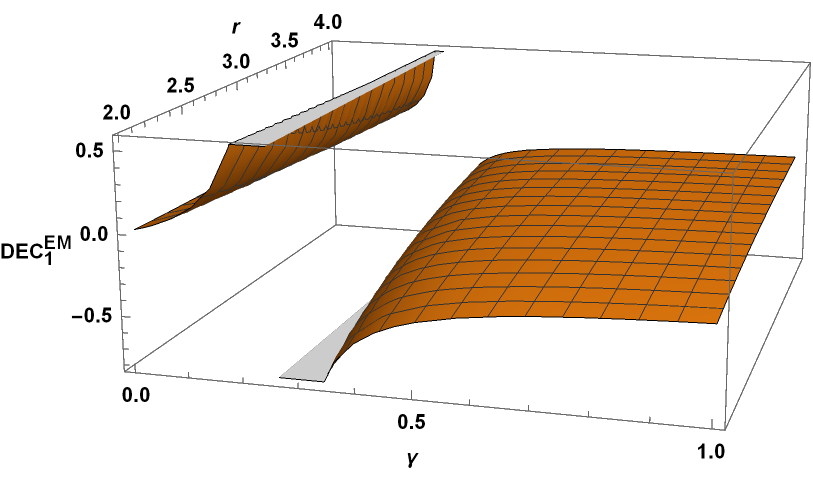}
    \caption{ ${\rm DEC}_{1}^{EM}$ for different values of parameter $\gamma$ and $r$ outside the horizon for a particular value of $m=1$ and $Q=0.5$ in the Simpson-Visser model.}
  \end{minipage}
  \end{figure}

 \begin{equation}\label{e2}
      {\rm SEC}_{3}^{\phi, EM}\Longleftrightarrow \rho^{\phi, EM}+p_{1}^{\phi, EM} +2p_{2}^{\phi, EM}\ge 0.
 \end{equation}
The ${\rm SEC}$ and ${\rm NEC}$ share a similar interpretation in terms of their geometric form. The ${\rm SEC}$ requires that timelike geodesic congruences converge in small neighborhoods of every point in spacetime.
\begin{equation}\label{e3}
      {\rm DEC}_{1,2}^{\phi, EM}\Longrightarrow \rho^{\phi, EM}-p_{1,2}^{\phi, EM}\ge 0,
 \end{equation}
The geometrical forms of the dominant energy condition $({\rm DEC})$ lack clarity in their interpretations. Nevertheless, the physical form of ${\rm DEC}$ can be understood as follows: any timelike observer will perceive the mass-energy density as positive or zero, and will also observe the energy-momentum flux as causal, flowing in the same direction as the observer's proper time. Finally, the other effective form of ${\rm DEC}$ and ${\rm WEC}$ is as follows
\begin{equation}\label{e4}
    {\rm DEC}_{3}^{\phi, EM}={\rm WEC}_{3}^{\phi, EM}\Longleftrightarrow \rho^{\phi, EM}\ge 0.
 \end{equation}

The scalar and electromagnetic fields are denoted by the indices $\phi$ and $EM$, respectively. The stress-energy tensor for the fluid quantities in the region where $f(r)>0$ is given by
\begin{equation}
    (S_{\,\,\nu}^{\mu})^{\phi, EM}={\rm diag}[\rho^{\phi, EM}, -p_{1}^{\phi, EM}, -p_{2}^{\phi, EM}, -p_{2}^{\phi, EM}].
\end{equation}
The scalar and electromagnetic parts of the stress-energy tensor are represented by $(S_{\,\,\nu}^{\mu})^{\phi}$ and $(S_{\,\,\nu}^{\mu})^{EM}$, respectively and their sum is denoted by $S^{\mu}_{\,\,\nu}$. The fluid quantities for the scalar part are expressed by

\begin{equation}\label{es1}
    \rho^{\phi}=\frac{-V+(1-2\gamma)f(r)(\phi^{\prime})^{2}}{4\gamma -1},~~~~~~~ p^{\phi}_{1}=\frac{V+(1-6\gamma)f(r)(\phi^{\prime})^{2}}{4\gamma -1},~~~~~~~ p^{\phi}_{2}=\frac{V-(1-2\gamma)f(r)(\phi^{\prime})^{2}}{4\gamma -1}.
\end{equation}

The fluid quantities for the electromagnetic field can be derived as follows: 
\begin{equation}\label{es4}
    \rho^{EM}=\frac{2\gamma Q^{2}L_{F}-\sigma^{4}L}{(4\gamma-1)\sigma^{4}},~~~~~~~~~ p^{EM}_{1}=\frac{-2\gamma Q^{2}L_{F}+\sigma^{4}L}{(4\gamma-1)\sigma^{4}},~~~~~~~~~~ p^{EM}_{2}=\frac{(2\gamma-1) Q^{2}L_{F}+\sigma^{4}L}{(4\gamma-1)\sigma^{4}}.
\end{equation}

Using the above considerations and equation $(\ref{e1})$, the energy conditions for the region where $f(r)>0$ can be obtained. The null, weak and strong energy conditions for scalar field are defined by
 \begin{eqnarray}\label{ec1}
    {\rm NEC}_{1}^{\phi}={\rm WEC}_{1}^{\phi}={\rm SEC}_{1}^{\phi}\Longleftrightarrow    -2f(r)(\phi^{\prime})^{2}  \ge 0, ~~~~~~~~~~~~~~~~~~~~~~~~~ {\rm NEC}_{2}^{\phi}={\rm WEC}_{2}^{\phi}={\rm SEC}_{2}^{\phi}\Longleftrightarrow  0,\\\nonumber
     {\rm SEC}_{3}^{\phi}\Longleftrightarrow \frac{2V-4\gamma f(r) (\phi^{\prime})^{2}}{4\gamma -1}  \ge 0,~~~~~~~~~~~~~~~~~~~~~~~~~~~ {\rm DEC}_{1}^{\phi} \Longrightarrow \frac{-2V+4\gamma f(r) (\phi^{\prime})^{2}}{4\gamma -1}  \ge 0,\\\nonumber
      {\rm DEC}_{2}^{\phi} \Longrightarrow \frac{-2V+(2-4\gamma )f(r) (\phi^{\prime})^{2}}{4\gamma -1}  \ge 0, ~~~~~~~~~~ {\rm DEC}_{3}^{\phi}={\rm WEC}_{3}^{\phi} \Longleftrightarrow \frac{-V+(1-2\gamma)f(r)(\phi^{\prime})^{2}}{4\gamma -1}  \ge 0.
 \end{eqnarray}
  \begin{figure}
  \centering
  \begin{minipage}[b]{0.49\textwidth}
    \includegraphics[width=\textwidth]{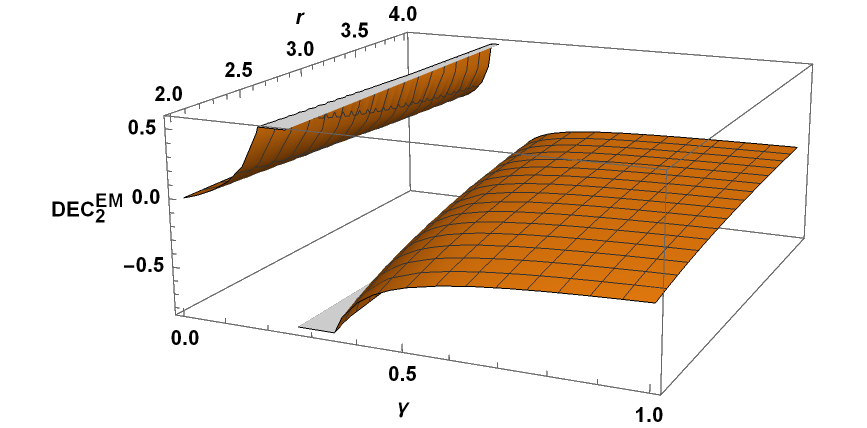}
    \caption{ ${\rm DEC}_{2}^{EM}$ for different values of parameter $\gamma$ and $r$ outside the horizon for a particular value of $m=1$ and $Q=0.5$ in the Simpson-Visser model.}
  \end{minipage}
  \hfill
  \begin{minipage}[b]{0.49\textwidth}
    \includegraphics[width=\textwidth]{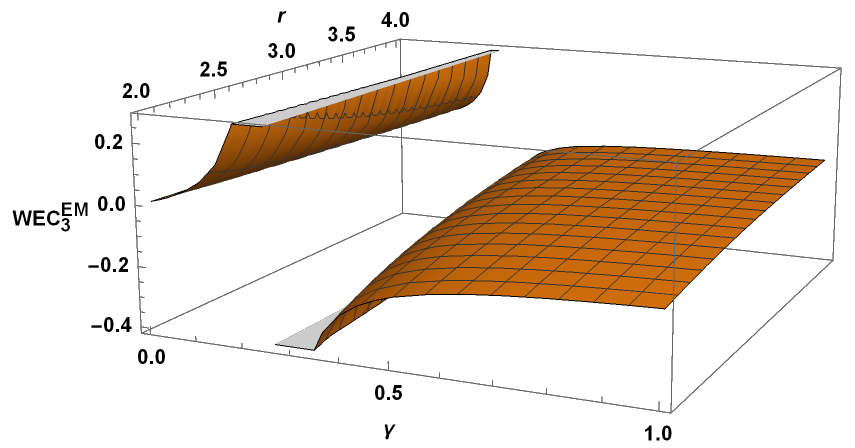}
    \caption{ ${\rm WEC}_{3}^{EM}$  for different values of parameter $\gamma$ and $r$ outside the horizon for a particular value of $m=1$ and $Q=0.5$ in the Simpson-Visser model.}
  \end{minipage}
\end{figure}

The energy conditions for the electromagnetic part in the region where $f(r)>0$ are derived from equations $(\ref{e1})$ and $(\ref{es4})$ and are expressed as follows:
\begin{eqnarray}\label{ec7}
     {\rm NEC}_{1}^{EM}={\rm WEC}_{1}^{EM}=SEC_{1}^{EM}\Longleftrightarrow      0,~~~~~~~~~~~~ {\rm NEC}_{2}^{EM}={\rm WEC}_{2}^{EM}={\rm SEC}_{2}^{EM}\Longleftrightarrow \frac{Q^{2}L_{F}}{\sigma^{4}} \ge 0, \\\nonumber
     {\rm SEC}_{3}^{EM}\Longleftrightarrow 2\frac{(2\gamma-1) Q^{2}L_{F}+\sigma^{4}L}{(4\gamma-1)\sigma^{4}} \ge 0,~~~~~~~~~~~~~~~~~~~~~~~~~~~~ {\rm DEC}_{1}^{EM} \Longrightarrow  \frac{4\gamma Q^{2}L_{F}-2\sigma^{4}L}{(4\gamma-1)\sigma^{4}} \ge 0,\\\nonumber
      {\rm DEC}_{2}^{EM} \Longrightarrow  \frac{ Q^{2}L_{F}-2\sigma^{4}L}{(4\gamma-1)\sigma^{4}} \ge 0,~~~~~~~~~~~~~~~~~~~~~~~~~ {\rm DEC}_{3}^{EM}={\rm WEC}_{3}^{EM} \Longleftrightarrow \frac{2\gamma Q^{2}L_{F}-\sigma^{4}L}{(4\gamma-1)\sigma^{4}} \ge 0.
 \end{eqnarray}

From equations $(\ref{ec1})$ and $(\ref{ec7})$, it follows that $ {\rm NEC}_{2}^{\phi}$ and ${\rm NEC}_{1}^{EM}$ are zero. The other energy conditions in Rastall gravity, except for $ {\rm NEC}_{1}^{\phi}$ and ${\rm NEC}_{2}^{EM}$, depend on the $\gamma$ parameter.
Now, we possess the necessary criteria to assess the energy conditions for the models examined in this study.
\begin{figure}[!tbp]
  \centering
  \begin{minipage}[b]{0.49\textwidth}
    \includegraphics[width=\textwidth]{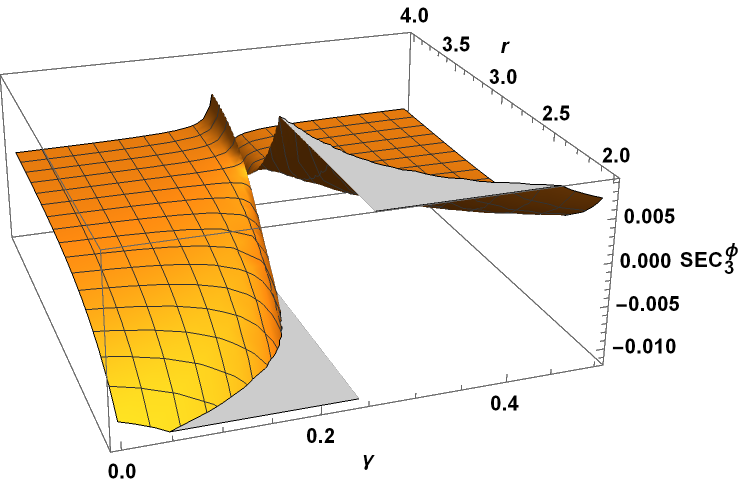}
    \caption{ ${\rm SEC}_{3}^{\phi}$  for different values of parameter $\gamma$ and $r$ outside the horizon for a particular value of $m=1$ and $Q=0.5$ in the Bardeen-Type model.}
  \end{minipage}
  \hfill
  \begin{minipage}[b]{0.49\textwidth}
    \includegraphics[width=\textwidth]{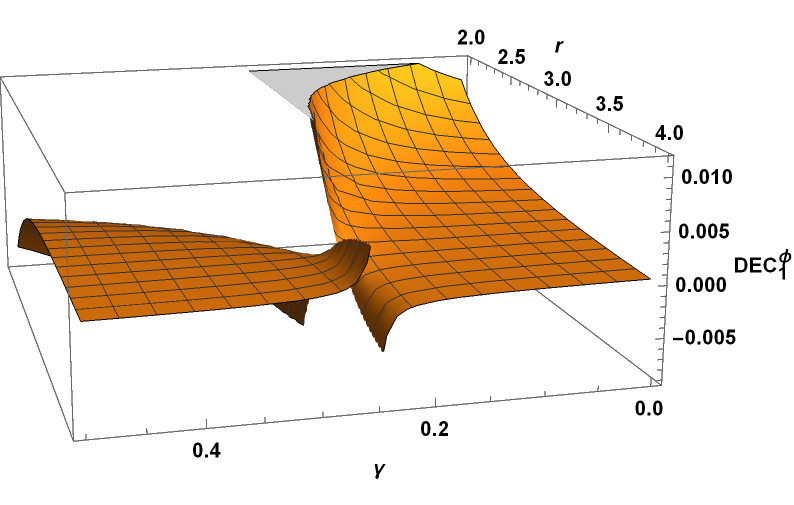}
    \caption{ ${\rm DEC}_{1}^{\phi}$  for different values of parameter $\gamma$ and $r$ outside the horizon for a particular value of $m=1$ and $Q=0.5$ in the Bardeen-Type model.}
  \end{minipage}
  \end{figure}

\subsection{Energy conditions for Simpson-Visser solution in Rastall gravity}
Equations $(\ref{phi})$ to $(\ref{LF})$  of the Simpson-Visser model are substituted into the energy conditions, leading to the following outcome. We begin by deriving the ${\rm NEC}_{1}^{\phi}$ for this model using equation $(\ref{simpsonsigma})$ and equation $(\ref{ec1})$. This gives us

\begin{equation}
   {\rm NEC}_{1}^{\phi} \Longleftrightarrow-2\frac{Q^{2}(\sqrt{r^2+Q^{2}}-2m)}{k^{2}(r^{2}+Q^{2})^{5/2}}\ge 0.
\end{equation}
 The violation of the ${\rm NEC}_{1}^{\phi}$ occurs in the region beyond the horizon where $f(r)>0$ (specifically, when $\sqrt{r^2+Q^{2}}>2m$). It is important to note that this energy condition for the scalar field remains unaffected by the Rastall parameter $\gamma$. Consequently, in Rastall gravity, the ${\rm NEC}_{1}^{\phi}$ holds the same significance as in general relativity for the Simpson-Visser black bounce solution. ${\rm SEC}_{3}^{\phi}$ is given by using equations $(\ref{simpsonsigma})$ and  $(\ref{ec1})$  
\begin{equation}
{\rm SEC}_{3}^{\phi}\Longleftrightarrow \frac{8mQ^{2}-20\gamma Q^{2}(\sqrt{r^2+Q^{2}}-2m)}{5k^{2}(4\gamma-1)(r^{2}+Q^{2})^{5/2}} \ge 0,
\end{equation}
when $\gamma$ equals zero, ${\rm SEC}_{3}^{\phi}$ is equivalent to ${\rm SEC}_{3}^{\phi}$ in general relativity for the Simpson-Visser solution. It is evident that this condition is violated in general relativity, but in Rastall gravity, it can be maintained due to the flexibility provided by the parameter $\gamma$. The energy condition is illustrated in Figure 7 for various values of $\gamma$, and it can be observed that there exists an interval where the energy condition is not violated. 
\begin{equation}
{\rm DEC}_{1}^{\phi}\Longrightarrow -\frac{8mQ^{2}-20\gamma Q^{2}(\sqrt{r^2+Q^{2}}-2m)}{5k^{2}(4\gamma-1)(r^{2}+Q^{2})^{5/2}} \ge 0.
\end{equation}
The parameter $\gamma$ in Rastall gravity allows for the possibility of satisfying the ${\rm DEC}_{1}^{\phi}$ for the Simpson-Visser black bounce solution. Figure $(8)$ shows how different values of $\gamma$ affect this energy condition for fixed values of $m=1$ and $Q=0.5$ 
\begin{figure}
  \centering
  \begin{minipage}[b]{0.49\textwidth}
    \includegraphics[width=\textwidth]{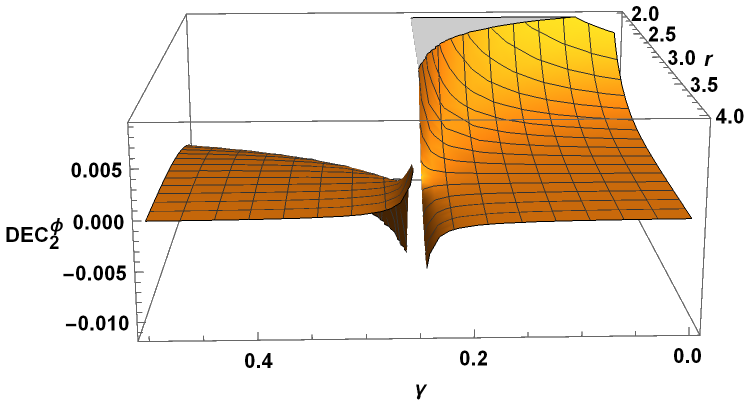}
    \caption{ ${\rm DEC}_{2}^{\phi}$  for different values of parameter $\gamma$ and $r$ outside the horizon for a particular value of $m=1$ and $Q=0.5$ in the Bardeen-Type model.}
  \end{minipage}
  \hfill
  \begin{minipage}[b]{0.49\textwidth}
    \includegraphics[width=\textwidth]{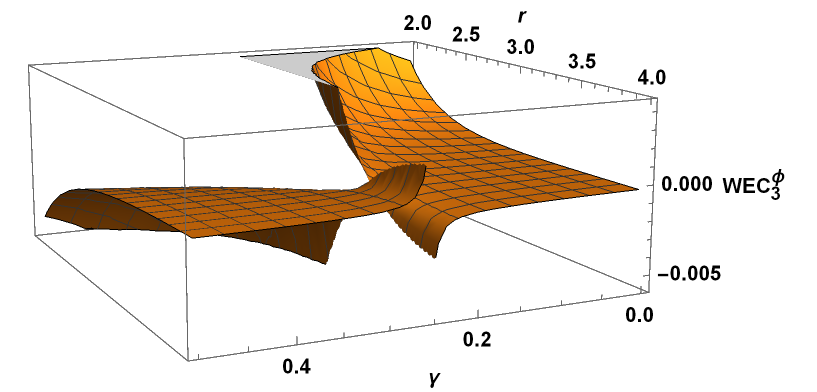}
    \caption{ ${\rm WEC}_{3}^{\phi}$  for different values of parameter $\gamma$ and $r$ outside the horizon for a particular value of $m=1$ and $Q=0.5$ in the Bardeen-Type model.}
  \end{minipage}
\end{figure}

\begin{equation}
{\rm DEC}_{2}^{\phi}\Longleftrightarrow \frac{-8mQ^{2}+5(2-4\gamma) Q^{2}(\sqrt{r^2+Q^{2}}-2m)}{5k^{2}(4\gamma-1)(r^{2}+Q^{2})^{5/2}} \ge 0.
\end{equation}
The ${\rm DEC}_{2}^{\phi}$ for the scalar field in Rastall gravity depends on the value of the parameter $\gamma$. Figure $(9)$ illustrates how this energy condition is satisfied for some values of $\gamma$ in a certain range.Finally, $ {\rm WEC}_{3}^{\phi}$ is given by using equations $(\ref{simpsonsigma})$ and $(\ref{ec1})$ as follows
\begin{equation}
    {\rm WEC}_{3}^{\phi} \Longleftrightarrow   \frac{-4mQ^{2}+5(1-2\gamma) Q^{2}(\sqrt{r^2+Q^{2}}-2m)}{5k^{2}(4\gamma-1)(r^{2}+Q^{2})^{5/2}} \ge 0.
 \end{equation}
Figure $(10)$ shows how the ${\rm WEC}_{3}^{\phi}$ varies with the parameter $\gamma$ for the scalar field in Rastall gravity. This energy condition reduces to the ${\rm WEC}_{3}^{\phi}$ for the Simpson-Visser solution in general relativity when $\gamma$ is zero. The ${\rm WEC}_{3}^{\phi}$ can be satisfied in both general relativity and Rastall gravity. Now, we proceed to derive the energy conditions for the electromagnetic field in the context of Rastall gravity for the Simpson-Visser black bounce solution. By using equations $(\ref{V})$, $(\ref{L})$ and $(\ref{ec7})$ we obtain ${\rm NEC}_{2}^{EM}$ as follows  

 \begin{equation}
     {\rm NEC}_{2}^{EM}\Longleftrightarrow \frac{3mQ^{2}}{k^{2}(Q^{2}+r^{2})^{5/2}} \ge 0.
 \end{equation}
 This energy condition does not depend on the Rastall parameter $\gamma$ and it coincides with the corresponding condition in general relativity. Therefore, ${\rm NEC}_{2}^{EM}$ is satisfied in general relativity and Rastall gravity. By applying equations $(\ref{L})$, $(\ref{LF})$ and $(\ref{ec7})$, ${\rm SEC}_{3}^{EM}$ is given by 
\begin{figure}[!tbp]
  \centering
  \begin{minipage}[b]{0.49\textwidth}
    \includegraphics[width=\textwidth]{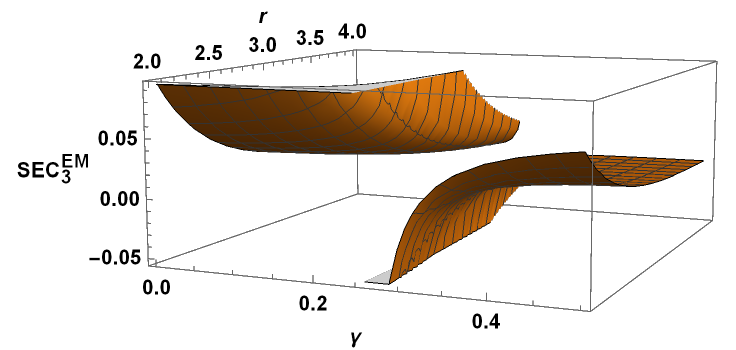}
    \caption{ ${\rm SEC}_{3}^{EM}$ for different values of parameter $\gamma$ and $r$ outside the horizon for a particular value of $m=1$ and $Q=0.5$ in the Bardeen-Type model.}
  \end{minipage}
  \hfill
  \begin{minipage}[b]{0.49\textwidth}
    \includegraphics[width=\textwidth]{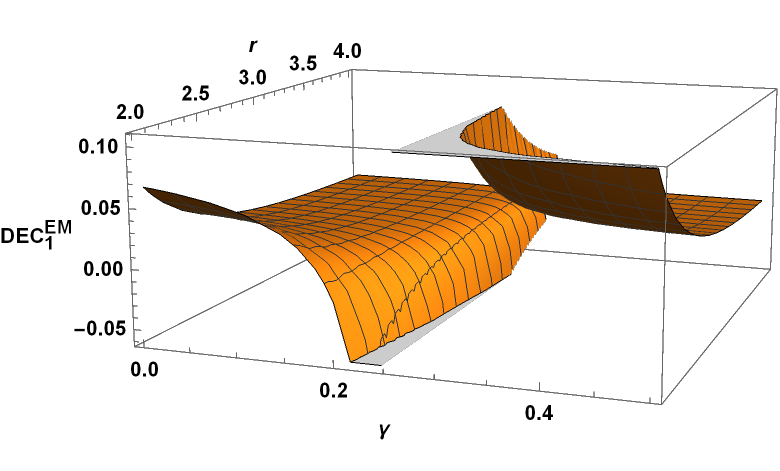}
    \caption{ ${\rm DEC}_{1}^{EM}$  for different values of parameter $\gamma$ and $r$ outside the horizon for a particular value of $m=1$ and $Q=0.5$ in the Bardeen-Type model.}
  \end{minipage}
  \end{figure}

\begin{equation}
     {\rm SEC}_{3}^{EM}\Longleftrightarrow \frac{-18mQ^{2}+20\gamma(r^{2}+Q^{2})^{3/2}}{5k^{2}(4\gamma-1)(r^{2}+Q^{2})^{5/2}} \ge 0,
 \end{equation}
 which depends on parameter $\gamma$. Figure $(11)$ shows how the ${\rm SEC}_{3}^{EM}$ for the electromagnetic field varies with the parameter $\gamma$ in the range $0<\gamma<1$. As seen in the figure, some values of $\gamma$ satisfy the ${\rm SEC}_{3}^{EM}$ for the electromagnetic field in Rastall gravity. We summarized the dominant energy conditions ${\rm DEC}_{1}^{EM}$ and ${\rm DEC}_{2}^{EM}$  as following relations 
 \begin{equation}
     {\rm DEC}_{1}^{EM} \Longrightarrow \frac{12(10\gamma -1)mQ^{2}-20\gamma(r^{2}+Q^{2})^{3/2}}{5k^{2}(4\gamma-1)(r^{2}+Q^{2})^{5/2}} \ge 0,~~~~~~~~~~~ {\rm DEC}_{2}^{EM} \Longrightarrow  \frac{3(20\gamma +1)mQ^{2}-20\gamma(r^{2}+Q^{2})^{3/2}}{5k^{2}(4\gamma-1)(r^{2}+Q^{2})^{5/2}} \ge 0,
 \end{equation}
  which are derived by using relations $(\ref{L})$, $(\ref{LF})$, and  $(\ref{ec7})$. Figures $(12)$ and $(13)$ demonstrate the presence of positive values for ${\rm DEC}_{1}^{EM}$ and ${\rm DEC}_{2}^{EM}$ at specific parameter $\gamma$ values. These findings provide evidence supporting the notion that these energy conditions in Rastall gravity may not be violated.Finally, $ {\rm WEC}_{3}^{\phi}$ is given by using equations $(\ref{L})$, $(\ref{LF})$ and $(\ref{ec7})$ as follows   
 \begin{equation}
    {\rm WEC}_{3}^{EM} \Longleftrightarrow \frac{6(10\gamma -1)mQ^{2}-10\gamma(r^{2}+Q^{2})^{3/2}}{5k^{2}(4\gamma-1)(r^{2}+Q^{2})^{5/2}} \ge 0
 \end{equation}
 In the context of Rastall gravity, the ${\rm WEC}_{3}^{EM}$ is depicted in figure $(14)$ as a function of the parameter $\gamma$ for the electromagnetic field. When $\gamma$ is zero, this energy condition is equivalent to the ${\rm WEC}_{3}^{EM}$ for the Simpson-Visser solution in general relativity. While the ${\rm WEC}_{3}^{EM}$ is always fulfilled in general relativity, Rastall gravity allows for the potential of non-violation with certain values of $\gamma$.

In the Simpson-Visser model of Rastall gravity, it has been observed that all energy conditions lead to the same condition as that of the GR model when the parameter $\gamma$  is set to zero. While the null condition ${\rm NEC}_{1}^{\phi}$ for the scalar field is violated outside the event horizon, it is possible to maintain other energy conditions for the scalar field for specific values of $\gamma$ and other parameters. In the case of an electromagnetic field, the null energy condition  ${\rm NEC}_{2}^{EM}$ is not violated outside the horizon, and it is feasible to choose $\gamma$  in a way that does not lead to the violation of any or some of the energy conditions. It is worth noting that the Rastall parameter $\gamma$  does not affect the energy conditions ${\rm NEC}_{1}^{\phi}$ and ${\rm NEC}_{2}^{EM}$, while the energy conditions  $ {\rm NEC}_{2}^{\phi}$ and ${\rm NEC}_{1}^{EM}$ are independent of$\gamma$  and vanish.

Figures $(7)$ to $(10)$ show the energy conditions for the scalar field $\phi$. We observe that ${\rm SEC}_{3}^{\phi}$ behaves differently from the other energy conditions ${\rm DEC}_{1}^{\phi}$ , ${\rm DEC}_{2}^{\phi}$ and ${\rm WEC}_{3}^{\phi}$ for $0<\gamma<0.5$ and $2<r<4$. These values of $r$ are chosen to represent the region outside the event horizon of the Simpson-Visser model, where $\sqrt{r^{2}+Q^{2}}>2m$. We set $m=1$ and $Q=0.5$ to ensure that the interval $2<r<4$ is valid. Figures $(8)$ to $(10)$ indicate that for Simpson-Visser solution in Rastall gravity, the energy conditions ${\rm DEC}_{1}^{\phi}$, ${\rm DEC}_{2}^{\phi}$ and ${\rm WEC}_{3}^{\phi}$ can be satisfied for some values of $\gamma$ and $r$ simultaneously.

Figures $(11)$ to $(14)$ display the energy conditions for the nonlinear electromagnetic fields. It is evident that ${\rm SEC}_{3}^{EM}$ exhibits distinct behavior compared to ${\rm DEC}_{1}^{EM}$, ${\rm DEC}_{2}^{EM}$, and ${\rm WEC}_{3}^{EM}$ when $0<\gamma<0.5$ and $2<r<4$. However, there exists a specific region in the figure where the energy condition remains preserved. To ensure the validity of the interval $2<r<4$, we once again assign $m=1$ and $Q=0.5$. Figures $(12)$ to $(14)$ demonstrate that it is possible to satisfy the energy conditions ${\rm DEC}_{1}^{EM}$, ${\rm DEC}_{2}^{EM}$, and ${\rm WEC}_{3}^{EM}$ simultaneously for certain values of $\gamma$ and $r$, for Simpson-Visser model in Rastall gravity. 

 \begin{figure}
  \centering
  \begin{minipage}[b]{0.49\textwidth}
    \includegraphics[width=\textwidth]{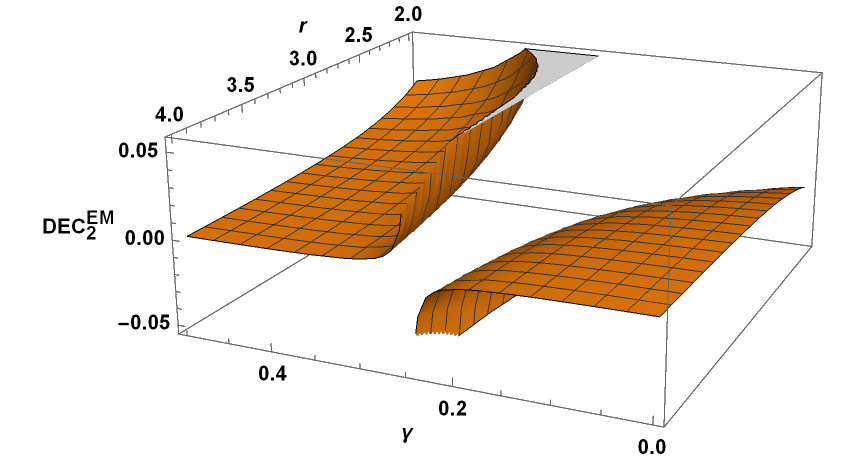}
    \caption{ ${\rm DEC}_{2}^{EM}$  for different values of parameter $\gamma$ and $r$ outside the horizon for a particular value of $m=1$ and $Q=0.5$ in the Bardeen-Type model.}
  \end{minipage}
  \hfill
  \begin{minipage}[b]{0.49\textwidth}
    \includegraphics[width=\textwidth]{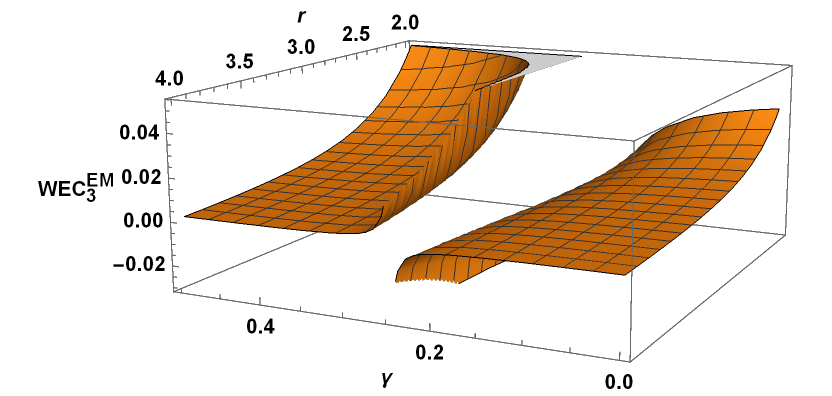}
    \caption{ ${\rm WEC}_{3}^{EM}$  for different values of parameter $\gamma$ and $r$ outside the horizon for a particular value of $m=1$ and $Q=0.5$ in the Bardeen-Type model.}
  \end{minipage}
\end{figure}

\subsection{Energy conditions for Bardeen-type solution in Rastall gravity} 

Energy conditions for a scalar field in the Bardeen-type black bounce can be obtained by inserting equation $(\ref{sigma2})$ into $(\ref{ec1})$. The resulting expression for ${\rm NEC}_{1}^{\phi}$ is given by:
 \begin{equation}
     {\rm NEC}_{1}^{\phi} \Longleftrightarrow    \frac{-2Q^{2}((r^{2}+Q^{2})^{3/2}-2mr^{2})}{k^{2}(r^{2}+Q^{2})^{7/2}}  \ge 0,
 \end{equation}
and it is not dependent on the parameter $\gamma$. The ${\rm NEC}_{1}^{\phi}$ is violated in the region outside the horizon where $f(r)>0$, which corresponds to the condition $(r^2+Q^{2})^{3/2}>2mr^{2}$. Therefore, the ${\rm NEC}_{1}^{\phi}$ has the same meaning in Rastall gravity as in general relativity for the Bardeen-type black bounce solution. The expression for ${\rm SEC}_{3}^{\phi}$ can be obtained by using equations $(\ref{sigma2})$ and $(\ref{ec1})$
\begin{equation}
     {\rm SEC}_{3}^{\phi}\Longleftrightarrow \frac{(56+280\gamma)mQ^{2}r^{2}-140\gamma Q^{2}(r^{2}+Q^{2})^{3/2}-64mQ^{4}}{35k^{2}(4\gamma-1)(r^{2}+Q^{2})^{7/2}}  \ge 0.
 \end{equation}
  
 When $\gamma$ equals zero, ${\rm SEC}_{3}^{\phi}$ is equivalent to ${\rm SEC}_{3}^{\phi}$ in general relativity for the Bardeen-type solution. This condition can be satisfied in general relativity, and in Rastall gravity, it can be maintained due to the flexibility provided by the parameter $\gamma$. The energy condition is illustrated in Figure $(15)$ for various values of $\gamma$, and it can be observed that there is a range of values of $\gamma$ that do not violate the energy condition
 \begin{equation}
     {\rm DEC}_{1}^{\phi} \Longrightarrow -\frac{(56+280\gamma)mQ^{2}r^{2}-140\gamma Q^{2}(r^{2}+Q^{2})^{3/2}-64mQ^{4}}{35k^{2}(4\gamma-1)(r^{2}+Q^{2})^{7/2}} \ge 0,
 \end{equation}
 the parameter $\gamma$ in Rastall gravity enables the satisfaction of the ${\rm DEC}_{1}^{\phi}$ for the Bardeen-type black bounce solution. Figure $(16)$ shows how different values of $\gamma$ affect the energy condition for fixed values of $m=1$ and $Q=0.5$.
  \begin{equation}
     {\rm DEC}_{2}^{\phi} \Longrightarrow \frac{(-196+280\gamma)mQ^{2}r^{2}+35(2-4\gamma) Q^{2}(r^{2}+Q^{2})^{3/2}+64mQ^{4}}{35k^{2}(4\gamma-1)(r^{2}+Q^{2})^{7/2}} \ge 0.
 \end{equation}
 The parameter $\gamma$ in Rastall gravity determines the validity of the ${\rm DEC}_{2}^{\phi}$ for the scalar field. Figure $(17)$ shows the range of values of $\gamma$ that satisfy this energy condition. The expression for ${\rm WEC}_{3}^{\phi}$ can be derived by using equations $(\ref{sigma2})$ and $(\ref{ec1})$ as follows:
 \begin{equation}
   {\rm WEC}_{3}^{\phi} =\frac{(-98+140\gamma)mQ^{2}r^{2}+35(1-2\gamma) Q^{2}(r^{2}+Q^{2})^{3/2}+32mQ^{4}}{35k^{2}(4\gamma-1)(r^{2}+Q^{2})^{7/2}}  \ge 0.
 \end{equation}
 
Figure $(18)$ demonstrates the dependence of the ${\rm WEC}_{3}^{\phi}$ for the scalar field on the parameter $\gamma$ in Rastall gravity. This energy condition coincides with the ${\rm WEC}_{3}^{\phi}$ for the Bardeen-type solution in general relativity when $\gamma$ is zero. The ${\rm WEC}_{3}^{\phi}$ can satisfy both general relativity and Rastall gravity. We continue by deriving the energy conditions for the electromagnetic field in Rastall gravity for the Bardeen-type black bounce solution. The expression for ${\rm NEC}_{2}^{EM}$ can be obtained by using equations $(\ref{V2})$, $(\ref{L2})$ and $(\ref{ec7})$ as follows:

  \begin{equation}
     {\rm NEC}_{2}^{EM} \Longleftrightarrow \frac{mQ^{2}(13r^{2}-2Q^{2})}{k^{2}(r^{2}+Q^{2})^{7/2}} \ge 0
 \end{equation}
 This energy condition is independent of the Rastall parameter $\gamma$ and has the same value as in general relativity. The ${\rm NEC}_{2}^{EM}$ can be satisfied in both general relativity and Rastall gravity for certain values of $Q$ and $m$. The expression for ${\rm SEC}_{3}^{EM}$ can be derived by using equations $(\ref{L2})$, $(\ref{LF2})$ and $(\ref{ec7})$ as follows:
\begin{equation}
     {\rm SEC}_{3}^{EM}\Longleftrightarrow \frac{(-546+1750\gamma)mQ^{2}r^{2}+(204-420\gamma)mQ^{4}}{35k^{2}(4\gamma-1)(r^{2}+Q^{2})^{7/2}} \ge 0,
 \end{equation}
  which depends on parameter $\gamma$. The variation of the ${\rm SEC}_{3}^{EM}$ for the electromagnetic field with respect to the parameter $\gamma$ in the interval $0<\gamma<0.5$ is depicted in Figure $(19)$. It can be observed from the figure that the ${\rm SEC}_{3}^{EM}$ for the electromagnetic field in Rastall gravity is satisfied for some values of $\gamma$. The dominant energy conditions ${\rm DEC}_{1}^{EM}$ and ${\rm DEC}_{2}^{EM}$ can be expressed as the following relations: 
 \begin{equation}
     {\rm DEC}_{1}^{EM} \Longrightarrow  \frac{(-364+1960\gamma)mQ^{2}r^{2}-64mQ^{4}}{35k^{2}(4\gamma-1)(r^{2}+Q^{2})^{7/2}} \ge 0,~~~~ {\rm DEC}_{2}^{EM} \Longrightarrow  \frac{(91+140\gamma)mQ^{2}r^{2}-(134+280\gamma)mQ^{4}}{35k^{2}(4\gamma-1)(r^{2}+Q^{2})^{7/2}} \ge 0,
 \end{equation}
  which are derived by using relations $(\ref{L2})$, $(\ref{LF2})$, and  $(\ref{ec7})$.The positive values of ${\rm DEC}_{1}^{EM}$ and ${\rm DEC}_{2}^{EM}$ for certain values of $\gamma$ are illustrated in Figures $(20)$ and $(21)$. These results indicate that the violation of these energy conditions in Rastall gravity may be avoided. The following expression for  ${\rm WEC}_{3}^{\phi}$ can be derived by using equations $(\ref{L})$, $(\ref{LF})$ and $(\ref{ec7})$: 
 \begin{equation}
    {\rm WEC}_{3}^{EM} \Longleftrightarrow \frac{(-184+980\gamma)mQ^{2}r^{2}-32mQ^{4}}{35k^{2}(4\gamma-1)(r^{2}+Q^{2})^{7/2}} \ge 0
 \end{equation}
Figure $(22)$ shows the variation of  ${\rm WEC}_{3}^{EM}$ for the electromagnetic field with respect to the parameter $\gamma$ in Rastall gravity. This energy condition reduces to the ${\rm WEC}_{3}^{EM}$ for the Bardeen-Type solution in general relativity when $\gamma$ is zero. Unlike general relativity, where the ${\rm WEC}_{3}^{EM}$ is always satisfied, Rastall gravity permits the possibility of non-violation for some values of $\gamma$.

 \begin{figure}
    \centering
    \includegraphics[width=0.5\linewidth]{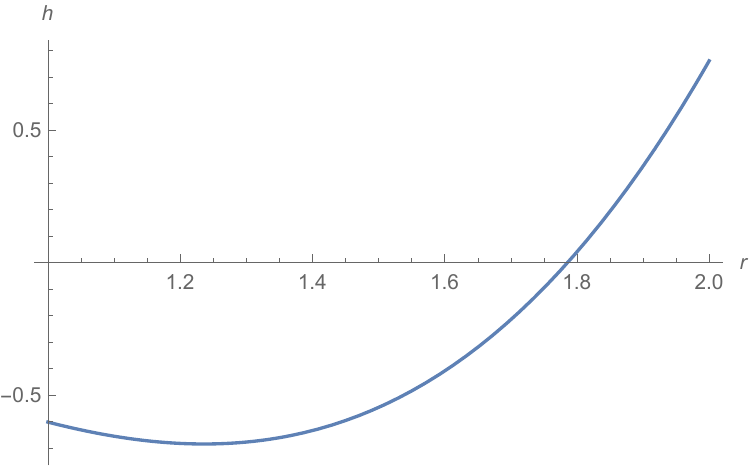}
    \caption{The behaviour of $h(r)=(r^{2}+Q^{2})^{3/2}-2mr^{2}$ with respect to $r$ for a particular value of $m=1$ and $Q=0.5$. }
    \label{fig:23}
\end{figure} 
The energy conditions of the GR model are recovered from the Bardeen-Type model of Rastall gravity when the parameter $\gamma$ is zero. The scalar field's null energy condition ${\rm NEC}_{1}^{\phi}$ is violated outside the event horizon, but other energy conditions for the scalar field can be satisfied by choosing appropriate values of $\gamma$ and other parameters. The electromagnetic field's null energy condition ${\rm NEC}_{2}^{EM}$ can be preserved outside the event horizon, and it is possible to select $\gamma$ such that some or all of the energy conditions are not violated. It is noteworthy that the energy conditions ${\rm NEC}_{1}^{\phi}$ and ${\rm NEC}_{2}^{EM}$ are unaffected by the Rastall parameter $\gamma$, while the energy conditions ${\rm NEC}_{2}^{\phi}$ and ${\rm NEC}_{1}^{EM}$ are zero and independent of $\gamma$.

Figures $(15)$ to $(18)$ display the energy conditions for the scalar field $\phi$. It is evident that ${\rm SEC}_{3}^{\phi}$ exhibits distinct behavior compared to the other energy conditions ${\rm DEC}_{1}^{\phi}$, ${\rm DEC}_{2}^{\phi}$, and ${\rm WEC}_{3}^{\phi}$ when $0<\gamma<0.5$ and $2<r<4$. These specific values of $r$ are selected to represent the region beyond the event horizon of the Bardeen-type solution, where $(r^{2}+Q^{2})^{3/2}>2mr^{2}$. As depicted in figure $(\ref{fig:23})$ and equation $(\ref{sigma2})$, by choosing $m=1$ and $Q=0.5$, one can confidently assert that the interval $2<r<4$ lies outside the event horizon. Figures $(16)$ to $(18)$ demonstrate that for the Bardeen-Type solution in Rastall gravity, the energy conditions ${\rm DEC}_{1}^{\phi}$, ${\rm DEC}_{2}^{\phi}$, and ${\rm WEC}_{3}^{\phi}$ can be simultaneously satisfied for certain values of $\gamma$ and $r$.

Figures $(19)$ to $(22)$ display the energy conditions of the nonlinear electromagnetic fields. It is noteworthy that the behavior of ${\rm SEC}_{3}^{EM}$ differs from that of the other energy conditions, namely ${\rm DEC}_{1}^{EM}$, ${\rm DEC}_{2}^{EM}$, and ${\rm WEC}_{3}^{EM}$, when $0<\gamma<0.5$ and $2<r<4$. Nevertheless, there exists a specific region within the figure where the energy condition remains preserved. To ensure the validity of the interval $2<r<4$, we have set $m=1$ and $Q=0.5$. Figures $(20)$ to $(22)$ demonstrate that the energy conditions ${\rm DEC}_{1}^{EM}$, ${\rm DEC}_{2}^{EM}$, and ${\rm WEC}_{3}^{EM}$ can be simultaneously satisfied for certain values of $\gamma$ and $r$ in the Bardeen-type model within the framework of Rastall gravity.

\section{Conclusion}

Our research delved into the black bounce solution in Rastall gravity, specifically examining the matter source required to describe this solution in a nonlinear electromagnetic field. We found that similar to general relativity, the use of a scalar or exotic matter is necessary to determine an appropriate matter source for the black bounce. In Rastall gravity, however, there is not enough flexibility to avoid the scalar field from being phantom when defining the source term solely with nonlinear electromagnetic fields. Using field equations, we developed a formalism to determine the material content of a general solution. Our findings led us to propose a theorem that links the need for the scalar field to be phantom with the violation of the null energy condition. We applied this method to the Simpson-Visser solution and the Bardeen-type black bounce solution in Rastall gravity and were able to determine the shape of $L(F)$, and the potential $V(\phi)$ that generate these solutions.
 Furthermore, we took into account the other energy conditions beyond the event horizon. It has been pointed out that in Rastall gravity, there exists greater adaptability to uphold the energy condition by adjusting the Rastall parameter $\gamma$ to some appropriate values. However, it should be noted that the null energy conditions for the scalar field and electromagnetic field become zero.

\section*{Acknowledgments}
The authors would like to thank an anonymous editor for useful and valuable comments which immensely helped us in improving the manuscript.

%%%%%%%%%%%%%%%%%%%%%%%%%%%%%%%%%%%%%%%%%%%%%%%%%%%%%%%
%%%%%%%%%%%%%%%%%%%%%%%%%%%%%%%%%%%%%%%%%%%%%%%%%

\end{document}